# Development of aging changes: self-accelerating and focalized

Jicun Wang-Michelitsch[1]*, Thomas M Michelitsch[2]

[1]Department of Medicine, Addenbrooke's Hospital, University of Cambridge, UK (Work address until 2007)

[2] Institut Jean le Rond d'Alembert (Paris 6), CNRS UMR 7190 Paris, France

## Abstract

Aging changes including age spots and atherosclerotic plaques develop in an inhomogeneous and accelerated manner. For understanding this phenomenon, some aging changes are analyzed by Misrepair mechanism, a mechanism proposed in Misrepair-accumulation theory. **I.** Misrepair is a strategy of repair for survival of an organism in situations of severe injuries; however a Misrepair alters the structure of a tissue, a cell or a molecule, which are the sub-structures of an organism. **II.** Misrepair of a sub-structure also alters the spatial relationship of this sub-structure with its neighbor sub-structures. Thus a Misrepair leads to increased damage-sensitivity and reduced repair-efficiency of these sub-structures. As a result, Misrepairs have a tendency to occur to the sub-structure and its neighbor sub-structures where an old Misrepair has taken place. In return, new Misrepairs will increase again the damage-sensitivity of these sub-structures and the surrounding sub-structures. By such a vicious circle, the frequency of Misrepairs to these sub-structures is increased and the range of affected sub-structures is enlarged after each time of Misrepair. Thus, accumulation of Misrepairs is focalized and self-accelerating. **III.** Focalized accumulation of Misrepairs leads to formation and growing of a "spot" or "plaque" in a tissue. Growing of a spot is self-accelerating, and old spots grow faster than new ones. New spots tend to develop close to old ones, resulting in an inhomogeneous distribution of spots. In conclusion, the inhomogeneous development of aging changes is a result of self-accelerated and focalized accumulation of Misrepairs; and the process of aging is self-accelerating.

Aging seems to be complex and mysterious, because it has a large variety of symptoms. An important issue in studying aging mechanism is to find out the common characteristics in different aging changes. We have observed that the age spots on skin and the atherosclerotic plaques in arterial walls develop in-homogeneously. The changing of age spots on number and on size seems to be somehow accelerated with time. It is interesting and important to know why these changes develop in this way and what the accelerating factor is. To interpret aging, we have proposed a Misrepair mechanism in Misrepair-accumulation theory (Wang, 2009). In our view, Misrepair mechanism is a surviving mechanism for an organism and for a species; however accumulation of Misrepairs result in aging of the organism. In the present paper, we take age spots as an example to analyze the developing characteristics of aging changes by Misrepair mechanism. Our discussion tackles the following issues:

   I.   Development of aging changes: inhomogeneous and accelerated

   II.  Aging as a result of accumulation of Misrepairs

   III. Effect of Misrepair on an organism: increased damage-sensitivity and reduced repair-efficiency of sub-structures

   IV.  Accumulation of Misrepairs: self-accelerating and focalized

   **V.** Conclusions

**I.   Development of aging changes: inhomogeneous and accelerated**

Age spots on the skin develop mainly on the face and on the back sides of two hands. Interestingly, age spots are always different to each other on size and on shape, and they distribute in-homogeneously. They have irregular shapes, and most of them are flat but some are protruding with deeper color. They are growing with time, and older ones are always bigger than younger ones. Smaller spots tend to develop in surrounding of a bigger one, which results in a satellite-like distribution of spots. Some spots in neighborhood can fuse together during growing (Figure 1A). Similarly, wrinkles on the face develop also in an inhomogeneous manner, and they have difference on lengths, depths, and directions. When new wrinkles appear successively, old wrinkles are growing in depth and in length (Figure 1B). Apart from age spots and wrinkles, the atherosclerotic plaques in arterial walls and the amyloid plaques in the brain have also inhomogeneous distributions (Figure 1C and 1D). Additionally, age spots seem to develop in an accelerated rate. The acceleration manifests on two aspects: one is the enlargement of each spot, and the other is the increase of number of spots. We can have such an experience: a visible age spot on skin looks "unchanged" for many years, but once it starts to grow, it does rapidly! Larger spots are often old spots, which grow faster than new ones. The number of spots increases faster and faster with time. For a person, the increase of number of spots in period of age 60-70 is much rapider than that in period of age 50-60.

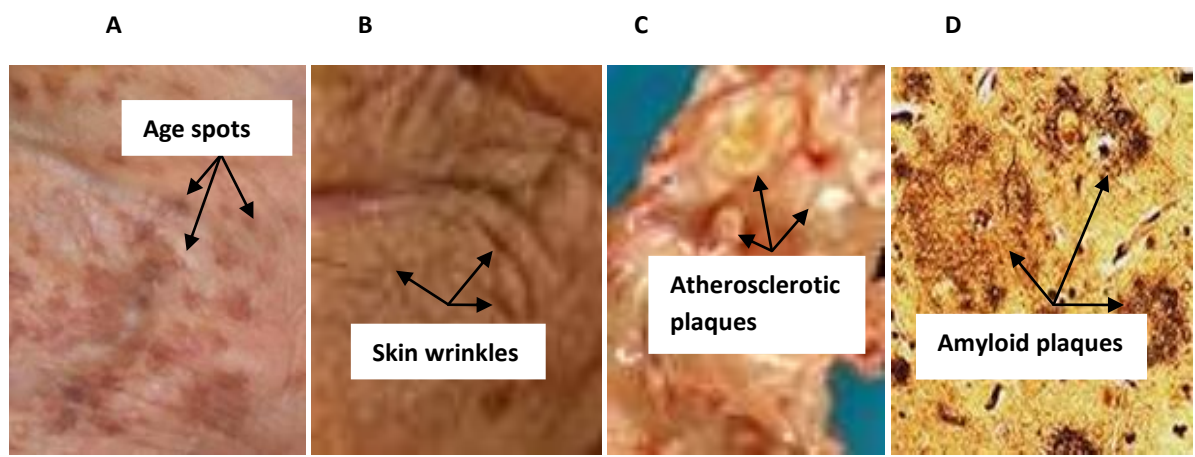

**Figure 1. Inhomogeneous development of aging changes**

Aging changes, including age spots **(A)**, skin wrinkles **(B)**, atherosclerosis plaques in arterial wall **(C)**, and amyloid plaques in brain **(D)**, are all inhomogeneous on distribution. In each type of changes, the spots are different to each other on size and shape (Picture C from Meddean.luc.edu and picture D from hoplive.com).

Some questions will be asked on age spots: Why do they develop in-homogeneously? Why do they grow continuously? What is the accelerating factor in the growing of spots? These questions are fundamental in understanding aging. Unfortunately neither damage (fault)-accumulation theory (Kirkwood, 2005) nor gene-controlling theory (Longo, 2005) can give satisfactory answer to them. On one hand, growing of an age spot cannot be a result of accumulation of damage. If a spot is produced by damage (injury), accumulation of random injuries should result in a homogenous distribution of spots. On the other hand, the in-homogeneous distribution of age spots cannot be a result of controlling of a gene. If a spot is produced by expression of a gene, the spots should have similar sizes and in a homogeneous distribution, because this gene should work in the same way in all the same types of cells. Thus, on interpreting the in-homogeneity of aging changes, these two theories are not tenable. In contrast, our novel theory, the Misrepair-accumulation theory, is exceptionally able to explain this in-homogeneity and answer the above questions.

## II. Aging as a result of accumulation of Misrepairs

To explain aging changes, we have proposed a generalized concept of Misrepair in Misrepair-accumulation theory (Wang, 2009). The new concept of Misrepair is defined as ***incorrect reconstruction of an injured living structure.*** This concept is applicable to all living structures including molecules (DNAs), cells, tissues, and organs. For example, scar formation is a Misrepair of epidermal tissue. When a complete repair is impossible to achieve for a severe injury, Misrepair, a repair with altered materials and in an altered remodeling, is a "SOS repair" that is essential for maintaining structural integrity and for the survival of an organism. However, a Misrepair results in alteration of the structure and reduction of the functionality of a living structure. The structure-alterations caused by Misrepairs are irreversible and irremovable; therefore they accumulate and "deform" gradually a living structure, appearing as aging of it. Thus, aging of an organism is a process of accumulation of Misrepairs. Misrepairs are essential for an organism to be able to survive till reproduction age; therefore Misrepair mechanism is essential for the survival of a species. Aging of individuals is a sacrifice for species' survival! Aging can take place on the levels of molecules, cells, and tissue, respectively; however aging of an organism takes place essentially on tissue level. A change of the spatial relationship of cells/extracellular matrix (ECMs) in a tissue is *essential and sufficient* for causing a decline of tissue functionality and body functionality. Aging of our body does not essentially require aging of cells and molecules. Aging of a tissue is often the cause but not the effect of aging of cells.

In our view, age spots, atherosclerotic plaques, and skin wrinkles are all results of accumulation of Misrepairs. **I.** Development of a flat age spot is a result of accumulation of aged basal cells, which contain lipofuscin bodies. Deposition of an aged cell is a Misrepair of a tissue, since a complete repair is to remove the aged cell and replace it with a new cell. **II.** Development of an atherosclerotic plaque is a result of altered remodeling (Misrepair) of endothelium, since infusion of lipids into sub-endothelium makes normal sealing of endothelium impossible to achieve. Accumulation of lipids and accumulation of Misrepairs of endothelium in part of an arterial wall result in formation of a plaque. **III.** Wrinkle formation is a result of repeated remodeling of skin derma with collagen fibers for replacing broken elastic fibers and other ECMs. When an elastic fiber is broken in an extended state, it will be replaced by a "longer" collagen fiber. Accumulation of "longer" collagen fibers makes the skin larger and stiffer. When a repairing collagen fiber in stiff skin is broken in a compressed state; it will be replaced by a "shorter" collagen fiber. The shorter collagen fibers will restrict the extension of longer fibers. Longer fibers have to rest permanently in a folding state, leading to formation of a permanent wrinkle. In these aging changes, Misrepairs tend to accumulate focally, resulting in formation of a spot or a plaque. But what is the mechanism for this focalized accumulation of Misrepairs? This is the question we will discuss in the next parts.

### III. Effect of Misrepair on an organism: increased damage-sensitivity and reduced repair-efficiency of sub-structures

Misrepair is a strategy of repair, and it is essential for a long survival of an organism. However, Misrepair of a sub-structure (called **Z**) of an organism does not only alter the structure of **Z** but also alters the spatial relationship between **Z** and its neighbor sub-structures. Molecules, cells, and tissues are all sub-structures of an organism. In the same time, molecules are sub-structures of a cell, and cells are sub-structures of a tissue. Substance-transportation and information-transmission can be disturbed by an alteration of the spatial relationship of local sub-structures. As a result, the Misrepaired sub-structure **Z** and its neighbor sub-structures will have reduced efficiency on functions of adaptation and repair. These sub-structures will have reduced ability to make adaptive responses to environment changes, thus they have increased damage-sensitivity. Because of increased damage-sensitivity and reduced repair-efficiency, these sub-structures will have increased risk for injuries and Misrepairs. Thus, Misrepairs have a tendency to occur to the sub-structure and its neighbor sub-structures where an old Misrepair has taken place. For a tissue, under a constant damage, the location where the first Misrepair takes place may be random. But the locations of the second Misrepair and the successive ones will not be any more random. New Misrepairs tend to take place next to or close to old ones. The locations of old Misrepairs are the centers of accumulation of new Misrepairs, since these areas are weak on functionality.

An age spot on the skin is pathologically a group of basal cells that contain lipofuscin inclusion bodies. Accumulation of lipofuscin bodies is a sign of aging of a cell. In a tissue, all cells need to communicate with their neighbor cells for functioning and for survival. Deposition of aged cells in a tissue is a kind of Misrepair of tissue. When an aged cell remains in a tissue, with reduced functionality, this cell will affect the functionality of its neighbor cells. **Firstly**, the repair-efficiency of local tissue will be reduced, since this aged cell can interrupt local substance-transportation. **Secondly**, neighbor cells cannot make efficient adaptive responses to changes of environment, and they become more sensitive to damage. Thus the neighbor cells of an aged cell will have increased risk for injuries and Misrepairs (Figure 2A). In this way, deposition of an aged cell acclerates the aging of its neighbor cells. Similarly, increased damage-sensitivity can be also seen in Misrepaired arterial walls. When part of the elastic membrane of arterial wall is injured, collagen fibers are often used for replacing the broken elastic fibers. Lack of elasticity, the replacing collagen fiber makes its neighbor elastic fibers have an increased risk to disrupt when they are being loaded (Figure 2B).

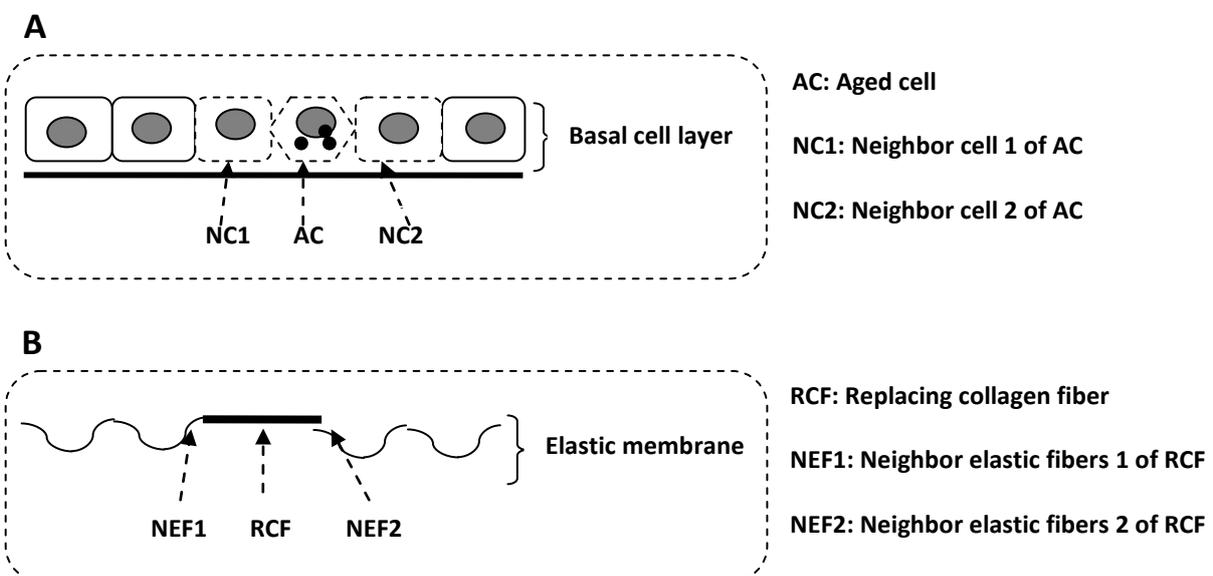

**Figure 2. Effect of Misrepair: increased damage-sensitivity and reduced repair-efficiency of sub-structures**

A Misrepair will alter the spatial relationship of a cell (or a molecule) and with its neighbor cells (or molecules). As a result, the neighbor cells (or molecules) will have increased damage-sensitivity and reduced repair-efficiency. For example, when an aged cell remains in a tissue, with reduced functionality, this cell will affect the functionality of its neighbor cells on repair and on adaptation. The neighbor cells will have an increased risk for injuries and Misrepairs. Thus deposition of an aged cell (**AC**) enhances the aging of its neighbor cells (**NC1** and **NC2**) (**A**). In part of an arterial wall, when elastic membrane is injured, collagen fibers are often used for replacing broken elastic fibers. Lack of elasticity, a replacing collagen fiber (**RCF**) will make its neighbor elastic fibers (**NEF1** and **NEF2**) have increased risk to disrupt when they are being loaded (**B**).

## IV. Accumulation of Misrepairs: self-accelerating and focalized

A Misrepair increases the risk of a sub-structure and its neighbor sub-structures for Misrepairs. New Misrepairs in these sub-structures will in return increase again the damage-sensitivity and the repair-efficiency of local sub-structures in a larger range. By such a vicious circle, the frequency of Misrepairs to these sub-structures will be increased and the range of affected sub-structures will be enlarged after each time of Misrepair (Figure 3). A Misrepair has a cascade amplifying effect, and an old Misrepair will promote new Misrepairs to take place to local sub-structures. Thus accumulation of Misrepairs is self-accelerating. For example, an aged cell in a tissue will enhance the aging of itself and the aging of its neighbor cells. In this way, more and more neighbor cells become aging and the group of aged cells is enlarged gradually. In part of an arterial wall, once an elastic fiber is broken and replaced by a collagen fiber, this replacement will promote disruption of neighbor elastic fibers and replacement of them by collagen fibers. In this way, the remodeling area of local arterial wall is enlarged gradually.

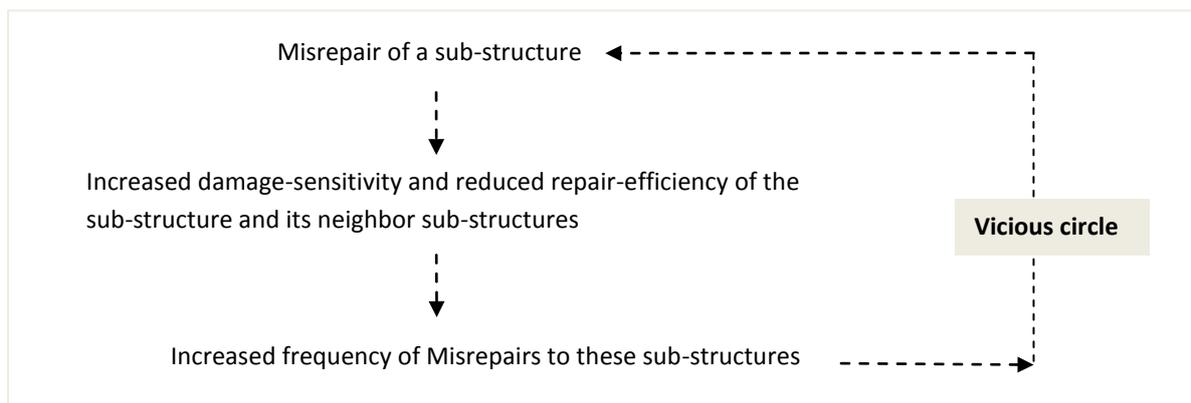

Figure 3. Accumulation of Misrepairs: a self-accelerating process

A Misrepair makes a sub-structure and its neighbor sub-structures have increased damage-sensitivity and reduced repair-efficiency. Thus Misrepairs have a tendency to occur to the sub-structure and its neighbor sub-structures where an old Misrepair has taken place. In return, new Misrepairs will further increase the damage-sensitivity and reduce the repair-efficiency of these sub-structures and the surrounding sub-structures. By such a vicious circle, the frequency of Misrepairs to these sub-structures will be increased and the range of affected sub-structures will be enlarged after each time of Misrepair **(Vicious circle).** Therefore, aging as a process of accumulation of Misrepairs is self-accelerating.

The tendency of new Misrepairs to occur to the locations of old ones also makes the accumulation of Misrepairs focalized and inhomogeneous. In a tissue, focalized accumulation of Misrepairs makes a Misrepaired area larger and larger, forming a visible "spot". As shown in Figure 4, in early stage (Stage 1), the earliest generation of Misrepairs (No. 1) takes place randomly. In stage 2, the No. 1 generation of Misrepairs promotes the occurrence of No. 2 generation of Misrepairs (No. 2), which accumulate to the locations of No. 1. In stage 3, the No. 2 generation of Misrepairs promotes the occurrence of No. 3 generation of Misrepairs (No. 3), which accumulate to the locations of No. 2. Aggregation of Misrepairs from different generations in a neighborhood results in formation of a "spot". The location of a spot in a tissue is determined by the location of the first Misrepair in this area. Further Misrepairs accumulate next to each other resulting in the "growing" of a spot with irregular shape. The new Misrepairs that take place far away from old ones will become centers of new spots. By the same mechanism, new spots tend to develop close to old ones, resulting in an in-homogenous distribution of spots. Bigger spots are always the older ones, because they have longer time of accumulation of Misrepairs.

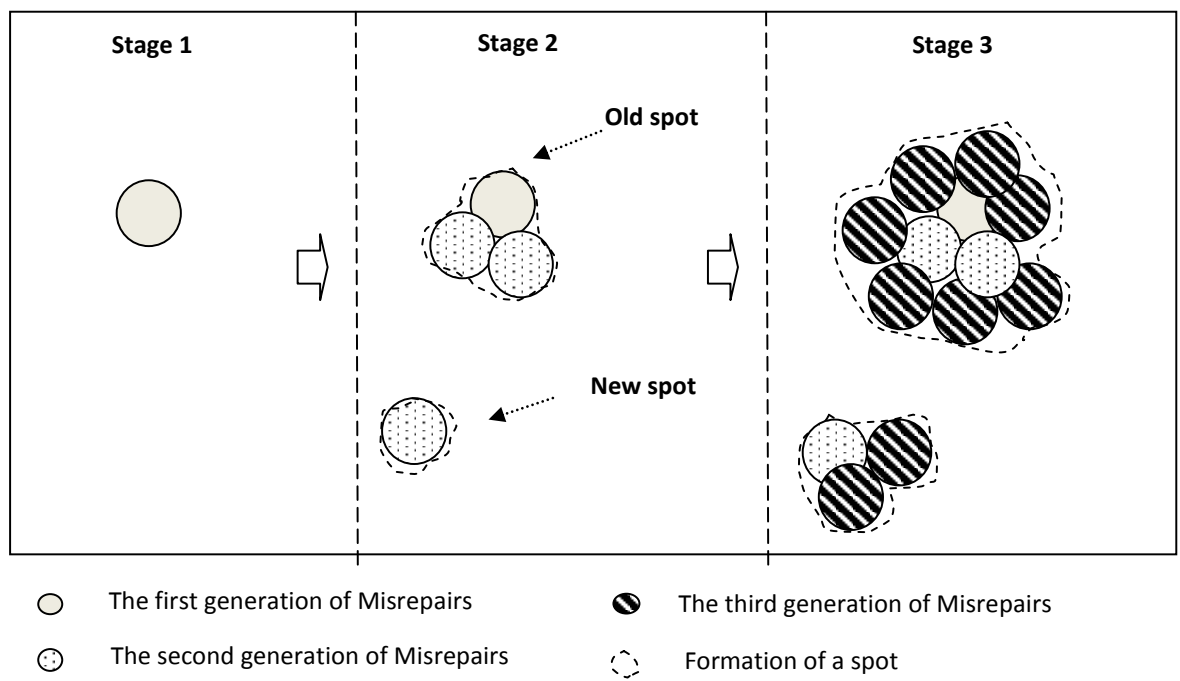

**Figure 4. Development of a spot: as a result of focalized accumulation of Misrepairs**

Development of a spot in a tissue is a result of focalized accumulation of Misrepairs of the tissue. In the beginning, the earliest generation of Misrepairs (**No. 1**) takes place randomly (**Stage 1**). Then, **No. 1** generation of Misrepairs promotes the occurrence of **No. 2** generation of Misrepairs (**No. 2**), which accumulate to the locations of No. 1 (**Stage 2**). Later, **No. 2** generation of Misrepairs promotes the occurrence of **No. 3** generation of Misrepairs (**No. 3**), which accumulate to the locations of No. 2 (**Stage 3**). Aggregation of Misrepairs from different generations in a neighborhood results in formation of a "spot". A spot developed in this way is irregular in shape and "grows" with time.

## V. Conclusions

Development of an aging change is a result of accumulation of Misrepairs. A Misrepair of a sub-structure increases the damage-sensitivity and reduces the repair-efficiency of the sub-structure and its neighbor sub-structures. As a result, Misrepairs have a tendency to occur to the sub-structure and its neighbor sub-structures where an old Misrepair has taken place. By a vicious circle, the frequency of Misrepairs to these sub-structures is increased and the range of affected sub-structures is enlarged after each time of Misrepair. Thus, accumulation of Misrepairs is focalized and self-accelerating. Focalized accumulation of Misrepairs leads to formation and growing of a "spot" or "plaque" in a tissue. Growing of a spot is self-accelerating. Old spots are always bigger than new ones, resulting in an inhomogeneous distribution of spots. In summary, the inhomogeneous distribution of aging changes is a result of centralized accumulation of Misrepairs; and aging is a self-accelerating process. Misrepair mechanism helps us understand why aging cannot be stopped.

*email : thomasjicun@gmail.com